\documentclass[a4paper,fleqn]{cas-dc}

\usepackage[numbers]{natbib}

\usepackage{bm}
\usepackage{braket}

\def\tsc#1{\csdef{#1}{\textsc{\lowercase{#1}}\xspace}}
\tsc{WGM}
\tsc{QE}

\usepackage[normalem]{ulem}  
\usepackage{color} 

\renewcommand\sout{\bgroup \color{red} \ULdepth=-.5ex \ULset}

\begin{document}
\let\WriteBookmarks\relax
\def\floatpagepagefraction{1}
\def\textpagefraction{.001}

\shorttitle{Compositeness of near-threshold states in charged hadronic systems}    

\shortauthors{T. Kinugawa, T. Hyodo}  

\title [mode = title]{Compositeness of near-threshold states in charged hadronic systems}  



%

\author[1]{Tomona Kinugawa}[orcid=0000-0002-8865-2753]

\cormark[1]


\ead{tomona.kinugawa@riken.jp}


\credit{Investigation, Methodology, Writing - original draft}

\affiliation[1]{organization={Nishina Center for Accelerator-Based Science, RIKEN},
            addressline={2-1 Hirosawa}, 
            city={Wako},
            postcode={351-0198}, 
            state={Saitama},
            country={Japan}}

\author[2]{Tetsuo Hyodo}[orcid=0000-0002-4145-9817]


\ead{hyodo@rcnp.osaka-u.ac.jp}


\credit{Investigation, Methodology, Writing - review \& editing}

\affiliation[2]{organization={Research Center for Nuclear Physics (RCNP), The University of Osaka},
            addressline={10-1 Mihogaoka}, 
            city={Ibaraki},
            postcode={567-0047}, 
            state={Osaka},
            country={Japan}}

\cortext[1]{Corresponding author}



\begin{abstract}
We quantify the internal structure of near-threshold bound, virtual, and resonance states in systems where Coulomb and short-range interactions coexist by evaluating the compositeness. Using the Coulomb-modified effective range expansion, we derive an expression for the compositeness in terms of the eigenenergy and Coulomb effective range in the weak-binding limit. We then apply the formulation to several near-threshold states in hadronic and nuclear systems, including $pp$, $\alpha\alpha$, $\Omega^{-}\Omega^{-}$, $\Omega_{ccc}^{++}\Omega_{ccc}^{++}$, $\Xi^{-}\alpha$, and $\Omega^{-}p$.
\end{abstract}





\maketitle

\section{Introduction}
\label{sec:intro}

Recent advances in accelerator experiments have led to the observation of numerous exotic hadrons~\cite{Hosaka:2016pey,Brambilla:2019esw}, which may possess internal structures beyond ordinary baryons and mesons composed of three quarks and a quark-antiquark pair. These states may take various forms, including compact multiquark states composed of four or more quarks, as well as hadronic molecules, in which mesons or baryons are weakly bound while preserving their individual degrees of freedom~\cite{Guo:2017jvc}. In particular, many candidates for exotic hadrons appear near two-body scattering thresholds. In this context, a precise investigation of near-threshold states provides an important key to unveiling their underlying structure.

To quantify the internal structure of an eigenstate $\ket{\Psi}$ from the viewpoint of the molecular component, we introduce the compositeness $X$, which represents the probability of finding a molecular component in the wavefunction~\cite{Weinberg:1965zz,Hyodo:2013nka,vanKolck:2022lqz,Kinugawa:2022fzn}. It is schematically expressed as
\begin{align}
\ket{\Psi} = \sqrt{X}\ket{\rm molecule} + \sqrt{Z}\ket{\rm non\ molecule}.
\end{align}
The quantity $Z$, called the elementarity, denotes the probability of finding non-molecular components in the wavefunction, such as multiquark states. The compositeness can be defined for general two-body systems, and it has been applied not only in hadron physics but also in nuclear and atomic systems~\cite{Kinugawa:2024crb}.

Near the two-body threshold, the low-energy universality emerges when the system is governed solely by short-range interactions. In this regime, the nature of near-threshold states is determined in a model-independent manner~\cite{Braaten:2004rn,Naidon:2016dpf}. For instance, in the limit where the binding energy approaches zero, the state becomes purely composite with $X = 1$~\cite{Hyodo:2014bda}. Furthermore, shallow bound states with small binding energies tend to be composite dominant, whereas near-threshold resonances with small excitation energies tend to have a non-composite structure~\cite{Hyodo:2013iga,Matuschek:2020gqe,Kinugawa:2023fbf,Kinugawa:2024kwb}.

When both scattering particles are charged, the Coulomb interaction acts in addition to the short-range interaction. 
Near-threshold states are also known to appear in the scattering of charged particles. A famous example is the $^{8}\mathrm{Be}$ nucleus, which is a resonance state in $\alpha\alpha$ scattering. If the Coulomb repulsion between the two $\alpha$ particles is artificially switched off, the $\alpha\alpha$ system would form a bound state due to the short-range interaction induced by the nuclear force. In reality, the Coulomb repulsion turns the bound state into a resonance slightly above the threshold. This demonstrates that the qualitative nature of a near-threshold state can change drastically due to the Coulomb interaction~\cite{Braaten:2004rn,Higa:2008dn}. A similar effect of the Coulomb interaction can be observed in hadronic systems as well as in exotic atoms where Coulomb attraction plays an essential role~\cite{Lyu:2021qsh,Hiyama:2022jqh}. Although the Coulomb interaction is in general weaker than the short-range force, its presence can significantly affect the properties of the near-threshold states. 

In this work, we analyze the internal structure of near-threshold states in systems with the Coulomb plus short-range interactions using the compositeness. Since low-energy universality is derived under the assumption of short-range interactions, its consequences are no longer applicable to systems where the Coulomb interaction is present. In the following, we quantitatively evaluate the compositeness of near-threshold eigenstates, taking specific two-body systems appearing in nuclear and hadron physics as examples.

\section{Compositeness of near-threshold states with Coulomb interaction}

Here, we calculate the compositeness of eigenstates in the presence of the Coulomb interaction. To this end, we employ an EFT formulation to derive the expression for the compositeness $X$ in terms of the energy derivative of the self-energy~\cite{Kinugawa:2024crb,Kinugawa:2026fob}. In the EFT framework, the low-energy scattering of charged particles is specified by the Coulomb scattering length $a_{s}^{C}$, the Coulomb effective range $r_{e}^{C}$, and $k_{C}=\mu\alpha Z_{1}Z_{2}$, where $\mu$ is the reduced mass of the system, $\alpha$ is the fine-structure constant, and $Z_{1}Z_{2}$ is the product of the charge numbers~\cite{Higa:2008dn}. The near-threshold eigenstate corresponds to a pole of the scattering amplitude. The eigenmomentum $k_{h}$ is determined by the pole condition as~\cite{Bethe:1949yr,Domcke:1983zz,Kong:1998sx,Kong:1999sf,Ando:2007fh,Higa:2008dn,Hammer:2008ra,Mochizuki:2024dbf}
\begin{align}
&-\frac{1}{a_{s}^{C}} + \frac{r_{e}^{C}}{2}k_{h}^{2} - ik_{h} \nonumber \\
&+ 2k_{C}\left \{\log\left(-i\frac{k_{h}}{|k_{C}|}\right) 
+ \psi\left(1 - \frac{ik_{C}}{k_{h}}\right)\right\} = 0,
\label{eq:pole-condition-ERE}
\end{align}
where $\psi$ is the digamma function. The eigenenergy is obtained as $E=k_{h}^{2}/(2\mu)$.

In this setup, the compositeness $X$ can be expressed in terms of the Coulomb effective range $r_{e}^{C}$ and the eigenmomentum $k_{h}$ as~\cite{Kinugawa:2026fob}:
\begin{align}
X &= \left[1 - \frac{r_{e}^{C}}{R^{C}}\right]^{-1}, 
\label{eq:wbr}
\\
R^{C} 
 &= -\sigma\frac{1}{k_{h}}\left[2i\left(\frac{k_{C}}{k_{h}}\right)^{2}\psi_{1}\left(i\frac{k_{C}}{k_{h}}\right)+ i - 2\left(\frac{k_{C}}{k_{h}}\right) \right], 
\end{align}
where $\sigma=-\text{sgn}(r_{e}^{C})$ and $\psi_{1}$ is the trigamma function. This is the weak-binding relation for the compositeness of near-threshold states in the presence of the Coulomb interaction.

For bound states in systems with a negative Coulomb effective range $r_{e}^{C}$, the compositeness $X$ defined in Eq.~\eqref{eq:wbr} takes values in the range $0 \leq X \leq 1$ and can therefore be interpreted as a probability. In contrast, for bound states in systems with $r_{e}^{C} > 0$, Eq.~\eqref{eq:wbr} yields $X>1$. Furthermore, for states with complex eigenenergies, $k_h$ becomes complex and hence the compositeness $X$ is also complex. In such cases, $X$ cannot be interpreted as a probability, and some prescription is required~\cite{Kinugawa:2024crb}. Here, following the method proposed in Ref.~\cite{Matuschek:2020gqe} for short-range interactions, we introduce
\begin{align}
X_{C} &= \sqrt{\frac{1}{1 + \left|-\left(\frac{r_{e}^{C}}{R^{C}}\right)^{2} + \frac{2r_{e}^{C}}{R^{C}}\right|}}.
\label{eq:X-C}
\end{align}
as a quantity that can be interpreted probabilistically. By construction, this definition yields $0\le X_{C}\le 1$ for arbitrary states, and hence it can be interpreted as a probability. 

\section{Application to near-threshold eigenstates}

Now we compute the compositeness of near-threshold states in the scattering of charged particles in hadron and nuclear physics. The properties of the scattering systems discussed in this work are summarized in Table~\ref{tab:examples}. As examples of systems with Coulomb repulsion, we consider the $pp$, $\alpha\alpha$, $\Omega^{-}\Omega^{-}$, and $\Omega_{ccc}^{++}\Omega_{ccc}^{++}$ systems. 
The existence of near-threshold states is suggested either from experimental data (virtual state in $pp$~\cite{Kong:1998sx}, ${}^{8}$Be nucleus in $\alpha\alpha$~\cite{Higa:2008dn}) or from lattice QCD calculations (bound state in $\Omega^{-}\Omega^{-}$~\cite{Gongyo:2017fjb}, resonance in $\Omega_{ccc}^{++}\Omega_{ccc}^{++}$~\cite{Lyu:2021qsh}). The values of the Coulomb scattering length and the Coulomb effective range are taken from the references listed in Table~\ref{tab:examples}.

For systems with Coulomb attraction, we examine the $\Xi^{-}\alpha$ and $\Omega^{-}p$ systems. In these systems, in addition to bound states originating from the Coulomb attraction (atomic states), the existence of bound states generated by the strong interaction has been suggested by theoretical analyses. The existence of a shallow bound state (the $\Xi$ hypernucleus $^{5}_{\Xi}$H) in the $\Xi^{-}\alpha$ system is predicted using the chiral EFT $\Xi N$ interaction~\cite{Le:2021gxa} and the HAL QCD $\Xi N$ interaction~\cite{Hiyama:2022jqh}. An $\Omega^{-}p$ dibaryon bound state is also found in lattice QCD calculations~\cite{HALQCD:2018qyu}.

\begin{table*}
 \caption{Summary of the scattering systems. $Z_{1}Z_{2}$ denotes the product of the charges of the scattering particles, $\mu$ the reduced mass, $a_{B}=1/|k_{C}|=1/(\mu \alpha |Z_1 Z_2|)$ the Bohr radius, $a_{s}^{C}$ the Coulomb scattering length, and $r_{e}^{C}$ the Coulomb effective range.\label{tab:examples}}
  \begin{tabular}{ccccccc}
    Scattering system & $Z_{1}Z_{2}$ & $\mu$ (MeV) & $a_{B}$ (fm) & $a_{s}^{C}$ (fm) & $r_{e}^{C}$ (fm) & Reference\\ \hline
    $pp$ & $+1$ & 469 & 57.6 & $-7.82$ & $2.83$ &\cite{ParticleDataGroup:2024cfk,Kong:1998sx}\\
    $\alpha\alpha$ & $+4$ & 1864 & 3.63 & $-1.80 \times 10^{3}$ & 1.083 & \cite{Mohr:2024kco,Higa:2008dn}\\ 
    $\Omega^{-}\Omega^{-}$ & $+1$ & 856 & 31.6 & $12.93$ & $1.21$ & \cite{Gongyo:2017fjb,byYanLyu-san}\\
    $\Omega_{ccc}^{++}\Omega_{ccc}^{++}$ & $+4$ & 2418.5 & 2.80 & $-19$ & $0.45$ & \cite{Lyu:2021qsh}\\ 
    $\Xi^{-} \alpha$ & $-2$ & 975.8 & 13.9 & $-12.6$ & $6.66$ & \cite{Hiyama:2022jqh,byKamiya-san}\\  
    $\Omega^{-} p$ & $-1$ & 613.0 & 44.1 & $3.2$ & $1.2$ & \cite{HALQCD:2018qyu,byYanLyu-san}\\ 
\hline
  \end{tabular}
\end{table*}

We first estimate the eigenmomentum $k_{h}$ from the Bohr radius $a_{B}$, Coulomb scattering length $a_{s}^{C}$ and Coulomb effective range $r_{e}^{C}$ in Table~\ref{tab:examples} using the pole condition~\eqref{eq:pole-condition-ERE}. We summarize the eigenenergy $E=k_{h}^{2}/(2\mu)$ in the first column of Table~\ref{tab:X}. From this analysis, we find that $pp$ scattering has a virtual-state pole on the second Riemann sheet of the complex energy plane. In contrast to the $nn$ system without the Coulomb interaction, the virtual pole in $pp$ scattering has a finite imaginary part due to the branch cut in the complex momentum plane. The $\alpha\alpha$ scattering has a resonance pole above the threshold, which is interpreted as the ground state of ${}^{8}$Be. 
The $\Omega^{-}\Omega^{-}$ system has a bound state below the threshold, while the $\Omega_{ccc}^{++}\Omega_{ccc}^{++}$ scattering has a resonance above the threshold. We find bound states in the $\Xi^{-}\alpha$ and $\Omega^{-}p$ systems with attractive Coulomb interaction, in accordance with the lattice QCD prediction. 
The estimated eigenenergies are found to be consistent with experimental data and lattice QCD result.

\begin{table*}
 \caption{
 Properties of the near-threshold eigenstates: the eigenenergy $E$, its classification, the compositeness $X$, and the interpretable compositeness $X_{C}$.
 \label{tab:X}}
  \begin{tabular}{ccccc}
    Scattering system & $E$ (MeV) & Classification & $X$ & $X_{C}$ \\ \hline 
    $pp$ & $-0.14 - 0.47i$ & Virtual state & $0.85 - 0.13i$ & $0.81$\\
    $\alpha\alpha$ & $0.083 - 8.3\times 10^{-7}i$ & Resonance (${}^{8}$Be) & $7.9 - 0.0013 i$ & $0.71$\\ 
    $\Omega^{-}\Omega^{-}$ & $-0.85$ & Bound state & $1.5$ & $0.81$\\
    $\Omega_{ccc}^{++}\Omega_{ccc}^{++}$ & $1.3 - 0.17i$ & Resonance & $1.5 - 0.2i$ & $0.79$ \\
    $\Xi^{-} \alpha$ & $-0.22$ & Bound state ($^{5}_{\Xi}$H) & $1.1$ & $1.0$\\
    $\Omega^{-} p$ & $-1.07$ & Bound state & $1.2$ & $0.87$\\
\hline
  \end{tabular}
\end{table*}

We then calculate the compositeness $X$ by Eq.~\eqref{eq:wbr} to quantify the structure of the states as summarized in Table~\ref{tab:X}. As mentioned above, for states with complex eigenenergies (the eigenstates in $pp$, $\alpha\alpha$, and $\Omega_{ccc}^{++}\Omega_{ccc}^{++}$), the compositeness $X$ becomes complex. On the other hand, for bound states in scattering systems with positive Coulomb effective ranges (the eigenstates in $\Omega^{-}\Omega^{-}$, $\Xi^{-}\alpha$, and $\Omega^{-}p$), $X$ is real but takes values larger than unity. Therefore, by evaluating the interpretable compositeness $X_{C}$ using Eq.~\eqref{eq:X-C}, we obtain $0 \leq X_{C} \leq 1$ for all states (see Table~\ref{tab:X}). This property allows us to interpret $X_{C}$ as the probability of the composite component of the eigenstate.

We find that all eigenstates examined in this work yield $X_{C}>0.5$ and are interpreted as composite dominant. In particular, our results indicate that $^{8}\mathrm{Be}$ exhibits a cluster-like structure, consistent with experimental observations and few-body calculations~\cite{Wiringa:2000gb,Otsuka:2022bcf}. The molecular-dominated structures obtained for the $\Omega^{-}\Omega^{-}$, $\Omega_{ccc}^{++}\Omega_{ccc}^{++}$, and $\Omega^{-}p$ dibaryons are in line with lattice QCD results showing that the inter-hadron distances are larger than the typical hadronic scale, indicating the molecular picture for these dibaryons.
These results demonstrate that the weak-binding relation in the presence of the Coulomb interaction provides a useful tool to characterize the internal structure of near-threshold states in nuclear and hadron physics.

\section{Summary}
\label{sec:summary}

We investigate the compositeness of near-threshold states in systems where Coulomb and short-range interactions coexist. 
By employing the Coulomb-modified effective range expansion, we determine the eigenenergies from the pole condition and clarify the nature of the corresponding eigenstates. Using the weak-binding relation for systems with the Coulomb interaction, we evaluate the compositeness of the eigenstates and introduce the interpretable compositeness $X_{C}$, which allows a probabilistic interpretation of the composite component.

Applying this framework to several systems in nuclear and hadron physics, including $pp$, $\alpha\alpha$, $\Omega^{-}\Omega^{-}$, $\Omega_{ccc}^{++}\Omega_{ccc}^{++}$, $\Xi^{-}\alpha$, and $\Omega^{-}p$, we quantitatively characterize the internal structure of the near-threshold states. The obtained results are consistent with experimental data, few-body calculations, and lattice QCD studies. These results demonstrate that the weak-binding relation in the presence of the Coulomb interaction provides a useful tool to study the structure of near-threshold states in nuclear and hadron physics. The present framework can be applied to various charged-particle systems and will be useful for future studies of exotic hadrons and hypernuclear states near thresholds.

\section*{Acknowledgments}
This work has been supported in part by the Grants-in-Aid for Scientific Research from JSPS (Grants
No.~JP23H05439 and 
No.~JP22K03637), 
    and by JST SPRING, Grant Number JPMJSP2156, by JST, the establishment of university fellowships towards the creation of science technology innovation, Grant No. JPMJFS2139,
    and by MIYAKO-MIRAI Project of Tokyo Metropolitan University.

\printcredits

\bibliographystyle{model1-num-names}

\bibliography{refs}

\begin{thebibliography}{34}
\expandafter\ifx\csname natexlab\endcsname\relax\def\natexlab#1{#1}\fi
\providecommand{\bibinfo}[2]{#2}
\ifx\xfnm\relax \def\xfnm[#1]{\unskip,\space#1}\fi
\bibitem[{Hosaka et~al.(2016)Hosaka, Iijima, Miyabayashi, Sakai, and
  Yasui}]{Hosaka:2016pey}
\bibinfo{author}{A.~Hosaka}, \bibinfo{author}{T.~Iijima},
  \bibinfo{author}{K.~Miyabayashi}, \bibinfo{author}{Y.~Sakai},
  \bibinfo{author}{S.~Yasui},
\newblock \bibinfo{title}{{Exotic hadrons with heavy flavors: $X$, $Y$, $Z$,
  and related states}},
\newblock \bibinfo{journal}{PTEP} \bibinfo{volume}{2016} (\bibinfo{year}{2016})
  \bibinfo{pages}{062C01}.
\bibitem[{Brambilla et~al.(2020)Brambilla, Eidelman, Hanhart, Nefediev, Shen,
  Thomas, Vairo, and Yuan}]{Brambilla:2019esw}
\bibinfo{author}{N.~Brambilla}, \bibinfo{author}{S.~Eidelman},
  \bibinfo{author}{C.~Hanhart}, \bibinfo{author}{A.~Nefediev},
  \bibinfo{author}{C.-P. Shen}, \bibinfo{author}{C.~E. Thomas},
  \bibinfo{author}{A.~Vairo}, \bibinfo{author}{C.-Z. Yuan},
\newblock \bibinfo{title}{{The $XYZ$ states: experimental and theoretical
  status and perspectives}},
\newblock \bibinfo{journal}{Phys. Rept.} \bibinfo{volume}{873}
  (\bibinfo{year}{2020}) \bibinfo{pages}{1--154}.
\bibitem[{Guo et~al.(2018)Guo, Hanhart, Mei\ss{}ner, Wang, Zhao, and
  Zou}]{Guo:2017jvc}
\bibinfo{author}{F.-K. Guo}, \bibinfo{author}{C.~Hanhart},
  \bibinfo{author}{U.-G. Mei\ss{}ner}, \bibinfo{author}{Q.~Wang},
  \bibinfo{author}{Q.~Zhao}, \bibinfo{author}{B.-S. Zou},
\newblock \bibinfo{title}{{Hadronic molecules}},
\newblock \bibinfo{journal}{Rev. Mod. Phys.} \bibinfo{volume}{90}
  (\bibinfo{year}{2018}) \bibinfo{pages}{015004}. \bibinfo{note}{[Erratum:
  Rev.Mod.Phys. 94, 029901 (2022)]}.
\bibitem[{Weinberg(1965)}]{Weinberg:1965zz}
\bibinfo{author}{S.~Weinberg},
\newblock \bibinfo{title}{{Evidence That the Deuteron Is Not an Elementary
  Particle}},
\newblock \bibinfo{journal}{Phys. Rev.} \bibinfo{volume}{137}
  (\bibinfo{year}{1965}) \bibinfo{pages}{B672--B678}.
\bibitem[{Hyodo(2013)}]{Hyodo:2013nka}
\bibinfo{author}{T.~Hyodo},
\newblock \bibinfo{title}{{Structure and compositeness of hadron resonances}},
\newblock \bibinfo{journal}{Int. J. Mod. Phys. A} \bibinfo{volume}{28}
  (\bibinfo{year}{2013}) \bibinfo{pages}{1330045}.
\bibitem[{van Kolck(2022)}]{vanKolck:2022lqz}
\bibinfo{author}{U.~van Kolck},
\newblock \bibinfo{title}{{Weinberg\textquoteright{}s Compositeness}},
\newblock \bibinfo{journal}{Symmetry} \bibinfo{volume}{14}
  (\bibinfo{year}{2022}) \bibinfo{pages}{1884}.
\bibitem[{Kinugawa and Hyodo(2022)}]{Kinugawa:2022fzn}
\bibinfo{author}{T.~Kinugawa}, \bibinfo{author}{T.~Hyodo},
\newblock \bibinfo{title}{{Structure of exotic hadrons by a weak-binding
  relation with finite-range correction}},
\newblock \bibinfo{journal}{Phys. Rev. C} \bibinfo{volume}{106}
  (\bibinfo{year}{2022}) \bibinfo{pages}{015205}.
\bibitem[{Kinugawa and Hyodo(2025)}]{Kinugawa:2024crb}
\bibinfo{author}{T.~Kinugawa}, \bibinfo{author}{T.~Hyodo},
\newblock \bibinfo{title}{{Compositeness of hadrons, nuclei, and atomic
  systems}},
\newblock \bibinfo{journal}{Eur. Phys. J. A} \bibinfo{volume}{61}
  (\bibinfo{year}{2025}) \bibinfo{pages}{154}.
\bibitem[{Braaten and Hammer(2006)}]{Braaten:2004rn}
\bibinfo{author}{E.~Braaten}, \bibinfo{author}{H.~W. Hammer},
\newblock \bibinfo{title}{{Universality in few-body systems with large
  scattering length}},
\newblock \bibinfo{journal}{Phys. Rept.} \bibinfo{volume}{428}
  (\bibinfo{year}{2006}) \bibinfo{pages}{259--390}.
\bibitem[{Naidon and Endo(2017)}]{Naidon:2016dpf}
\bibinfo{author}{P.~Naidon}, \bibinfo{author}{S.~Endo},
\newblock \bibinfo{title}{{Efimov Physics: a review}},
\newblock \bibinfo{journal}{Rept. Prog. Phys.} \bibinfo{volume}{80}
  (\bibinfo{year}{2017}) \bibinfo{pages}{056001}.
\bibitem[{Hyodo(2014)}]{Hyodo:2014bda}
\bibinfo{author}{T.~Hyodo},
\newblock \bibinfo{title}{{Hadron mass scaling near the s-wave threshold}},
\newblock \bibinfo{journal}{Phys. Rev. C} \bibinfo{volume}{90}
  (\bibinfo{year}{2014}) \bibinfo{pages}{055208}.
\bibitem[{Hyodo(2013)}]{Hyodo:2013iga}
\bibinfo{author}{T.~Hyodo},
\newblock \bibinfo{title}{{Structure of Near-Threshold s-Wave Resonances}},
\newblock \bibinfo{journal}{Phys. Rev. Lett.} \bibinfo{volume}{111}
  (\bibinfo{year}{2013}) \bibinfo{pages}{132002}.
\bibitem[{Matuschek et~al.(2021)Matuschek, Baru, Guo, and
  Hanhart}]{Matuschek:2020gqe}
\bibinfo{author}{I.~Matuschek}, \bibinfo{author}{V.~Baru},
  \bibinfo{author}{F.-K. Guo}, \bibinfo{author}{C.~Hanhart},
\newblock \bibinfo{title}{{On the nature of near-threshold bound and virtual
  states}},
\newblock \bibinfo{journal}{Eur. Phys. J. A} \bibinfo{volume}{57}
  (\bibinfo{year}{2021}) \bibinfo{pages}{101}.
\bibitem[{Kinugawa and Hyodo(2024{\natexlab{a}})}]{Kinugawa:2023fbf}
\bibinfo{author}{T.~Kinugawa}, \bibinfo{author}{T.~Hyodo},
\newblock \bibinfo{title}{{Compositeness of Tcc and X(3872) by considering
  decay and coupled-channels effects}},
\newblock \bibinfo{journal}{Phys. Rev. C} \bibinfo{volume}{109}
  (\bibinfo{year}{2024}{\natexlab{a}}) \bibinfo{pages}{045205}.
\bibitem[{Kinugawa and Hyodo(2024{\natexlab{b}})}]{Kinugawa:2024kwb}
\bibinfo{author}{T.~Kinugawa}, \bibinfo{author}{T.~Hyodo},
\newblock \bibinfo{title}{{Compositeness of near-threshold $s$-wave
  resonances}}  (\bibinfo{year}{2024}{\natexlab{b}}).
\bibitem[{Higa et~al.(2008)Higa, Hammer, and van Kolck}]{Higa:2008dn}
\bibinfo{author}{R.~Higa}, \bibinfo{author}{H.~W. Hammer},
  \bibinfo{author}{U.~van Kolck},
\newblock \bibinfo{title}{{alpha alpha Scattering in Halo Effective Field
  Theory}},
\newblock \bibinfo{journal}{Nucl. Phys. A} \bibinfo{volume}{809}
  (\bibinfo{year}{2008}) \bibinfo{pages}{171--188}.
\bibitem[{Lyu et~al.(2021)Lyu, Tong, Sugiura, Aoki, Doi, Hatsuda, Meng, and
  Miyamoto}]{Lyu:2021qsh}
\bibinfo{author}{Y.~Lyu}, \bibinfo{author}{H.~Tong},
  \bibinfo{author}{T.~Sugiura}, \bibinfo{author}{S.~Aoki},
  \bibinfo{author}{T.~Doi}, \bibinfo{author}{T.~Hatsuda},
  \bibinfo{author}{J.~Meng}, \bibinfo{author}{T.~Miyamoto},
\newblock \bibinfo{title}{{Dibaryon with Highest Charm Number near Unitarity
  from Lattice QCD}},
\newblock \bibinfo{journal}{Phys. Rev. Lett.} \bibinfo{volume}{127}
  (\bibinfo{year}{2021}) \bibinfo{pages}{072003}.
\bibitem[{Hiyama et~al.(2022)Hiyama, Isaka, Doi, and Hatsuda}]{Hiyama:2022jqh}
\bibinfo{author}{E.~Hiyama}, \bibinfo{author}{M.~Isaka},
  \bibinfo{author}{T.~Doi}, \bibinfo{author}{T.~Hatsuda},
\newblock \bibinfo{title}{{Probing the \ensuremath{\Xi}N interaction through
  inversion of spin-doublets in
  \ensuremath{\Xi}N\ensuremath{\alpha}\ensuremath{\alpha} nuclei}},
\newblock \bibinfo{journal}{Phys. Rev. C} \bibinfo{volume}{106}
  (\bibinfo{year}{2022}) \bibinfo{pages}{064318}.
\bibitem[{Kinugawa and Hyodo(2026)}]{Kinugawa:2026fob}
\bibinfo{author}{T.~Kinugawa}, \bibinfo{author}{T.~Hyodo},
\newblock \bibinfo{title}{{Compositeness of near-threshold eigenstates with
  Coulomb plus short-range interactions}},
\newblock \bibinfo{journal}{arXiv:2604.17813}  
	(\bibinfo{year}{2026}).
\bibitem[{Bethe(1949)}]{Bethe:1949yr}
\bibinfo{author}{H.~A. Bethe},
\newblock \bibinfo{title}{{Theory of the Effective Range in Nuclear
  Scattering}},
\newblock \bibinfo{journal}{Phys. Rev.} \bibinfo{volume}{76}
  (\bibinfo{year}{1949}) \bibinfo{pages}{38--50}.
\bibitem[{Domcke(1983)}]{Domcke:1983zz}
\bibinfo{author}{W.~Domcke},
\newblock \bibinfo{title}{{Analytic theory of resonances and bound states near
  Coulomb thresholds}},
\newblock \bibinfo{journal}{J. Phys. B: Atom. Mol. Phys.} \bibinfo{volume}{16}
  (\bibinfo{year}{1983}).
\bibitem[{Kong and Ravndal(1999)}]{Kong:1998sx}
\bibinfo{author}{X.~Kong}, \bibinfo{author}{F.~Ravndal},
\newblock \bibinfo{title}{{Proton proton scattering lengths from effective
  field theory}},
\newblock \bibinfo{journal}{Phys. Lett. B} \bibinfo{volume}{450}
  (\bibinfo{year}{1999}) \bibinfo{pages}{320--324}. \bibinfo{note}{[Erratum:
  Phys.Lett.B 458, 565--565 (1999)]}.
\bibitem[{Kong and Ravndal(2000)}]{Kong:1999sf}
\bibinfo{author}{X.~Kong}, \bibinfo{author}{F.~Ravndal},
\newblock \bibinfo{title}{{Coulomb effects in low-energy proton proton
  scattering}},
\newblock \bibinfo{journal}{Nucl. Phys. A} \bibinfo{volume}{665}
  (\bibinfo{year}{2000}) \bibinfo{pages}{137--163}.
\bibitem[{Ando et~al.(2007)Ando, Shin, Hyun, and Hong}]{Ando:2007fh}
\bibinfo{author}{S.-i. Ando}, \bibinfo{author}{J.~W. Shin},
  \bibinfo{author}{C.~H. Hyun}, \bibinfo{author}{S.~W. Hong},
\newblock \bibinfo{title}{{Low energy proton-proton scattering in effective
  field theory}},
\newblock \bibinfo{journal}{Phys. Rev. C} \bibinfo{volume}{76}
  (\bibinfo{year}{2007}) \bibinfo{pages}{064001}.
\bibitem[{Hammer and Higa(2008)}]{Hammer:2008ra}
\bibinfo{author}{H.~W. Hammer}, \bibinfo{author}{R.~Higa},
\newblock \bibinfo{title}{{A Model Study of Discrete Scale Invariance and
  Long-Range Interactions}},
\newblock \bibinfo{journal}{Eur. Phys. J. A} \bibinfo{volume}{37}
  (\bibinfo{year}{2008}) \bibinfo{pages}{193--200}.
\bibitem[{Mochizuki and Nishida(2024)}]{Mochizuki:2024dbf}
\bibinfo{author}{S.~Mochizuki}, \bibinfo{author}{Y.~Nishida},
\newblock \bibinfo{title}{{Universal bound states and resonances with Coulomb
  plus short-range potentials}},
\newblock \bibinfo{journal}{Phys. Rev. C} \bibinfo{volume}{110}
  (\bibinfo{year}{2024}) \bibinfo{pages}{064001}.
\bibitem[{Gongyo et~al.(2018)}]{Gongyo:2017fjb}
\bibinfo{author}{S.~Gongyo}, et~al.,
\newblock \bibinfo{title}{{Most Strange Dibaryon from Lattice QCD}},
\newblock \bibinfo{journal}{Phys. Rev. Lett.} \bibinfo{volume}{120}
  (\bibinfo{year}{2018}) \bibinfo{pages}{212001}.
\bibitem[{Iritani et~al.(2019)}]{HALQCD:2018qyu}
\bibinfo{author}{T.~Iritani}, et~al.,
\newblock \bibinfo{title}{{$N\Omega$ dibaryon from lattice QCD near the
  physical point}},
\newblock \bibinfo{journal}{Phys. Lett. B} \bibinfo{volume}{792}
  (\bibinfo{year}{2019}) \bibinfo{pages}{284--289}.
\bibitem[{Navas et~al.(2024)}]{ParticleDataGroup:2024cfk}
\bibinfo{author}{S.~Navas}, et~al.,
\newblock \bibinfo{title}{{Review of particle physics}},
\newblock \bibinfo{journal}{Phys. Rev. D} \bibinfo{volume}{110}
  (\bibinfo{year}{2024}) \bibinfo{pages}{030001}.
\bibitem[{Mohr et~al.(2025)Mohr, Newell, Taylor, and Tiesinga}]{Mohr:2024kco}
\bibinfo{author}{P.~J. Mohr}, \bibinfo{author}{D.~B. Newell},
  \bibinfo{author}{B.~N. Taylor}, \bibinfo{author}{E.~Tiesinga},
\newblock \bibinfo{title}{{CODATA recommended values of the fundamental
  physical constants: 2022*}},
\newblock \bibinfo{journal}{Rev. Mod. Phys.} \bibinfo{volume}{97}
  (\bibinfo{year}{2025}) \bibinfo{pages}{025002}.
\bibitem[{Lyu(????)}]{byYanLyu-san}
\bibinfo{author}{Y.~Lyu},
\newblock \bibinfo{journal}{private communication}.
\bibitem[{Kamiya(????)}]{byKamiya-san}
\bibinfo{author}{Y.~Kamiya},
\newblock \bibinfo{journal}{private communication}.
\bibitem[{Wiringa et~al.(2000)Wiringa, Pieper, Carlson, and
  Pandharipande}]{Wiringa:2000gb}
\bibinfo{author}{R.~B. Wiringa}, \bibinfo{author}{S.~C. Pieper},
  \bibinfo{author}{J.~Carlson}, \bibinfo{author}{V.~R. Pandharipande},
\newblock \bibinfo{title}{{Quantum Monte Carlo calculations of A = 8 nuclei}},
\newblock \bibinfo{journal}{Phys. Rev. C} \bibinfo{volume}{62}
  (\bibinfo{year}{2000}) \bibinfo{pages}{014001}.
\bibitem[{Otsuka et~al.(2022)Otsuka, Abe, Yoshida, Tsunoda, Shimizu, Itagaki,
  Utsuno, Vary, Maris, and Ueno}]{Otsuka:2022bcf}
\bibinfo{author}{T.~Otsuka}, \bibinfo{author}{T.~Abe},
  \bibinfo{author}{T.~Yoshida}, \bibinfo{author}{Y.~Tsunoda},
  \bibinfo{author}{N.~Shimizu}, \bibinfo{author}{N.~Itagaki},
  \bibinfo{author}{Y.~Utsuno}, \bibinfo{author}{J.~Vary},
  \bibinfo{author}{P.~Maris}, \bibinfo{author}{H.~Ueno},
\newblock \bibinfo{title}{{{\ensuremath{\alpha}}-Clustering in atomic nuclei
  from first principles with statistical learning and the Hoyle state
  character}},
\newblock \bibinfo{journal}{Nature Commun.} \bibinfo{volume}{13}
  (\bibinfo{year}{2022}) \bibinfo{pages}{2234}.

\end{thebibliography}



\end{document}